\documentclass[final,5p,times,twocolumn,nofootinbib,nopreprintline,sort&compress]{elsarticle}
\usepackage{amsmath,amssymb,amsfonts}
\usepackage{graphicx}
\usepackage{hyperref}
\usepackage{slashed}
\usepackage{booktabs,tabulary}
\usepackage{mciteplus}
\usepackage{xcolor}
\usepackage{soul}

\usepackage{bibentry}
\newcommand{\ignore}[1]{}
\newcommand{\nobibentry}[1]{{\let\nocite\ignore\bibentry{#1}}}

\makeatletter
\def\bibinfo@X@title#1,{\ignorespaces}
\makeatother

\begin{document}

\begin{frontmatter}

\title{Analytical dispersive parameterization for elastic scattering of spinless particles}
\author[Mainz]{Igor Danilkin}
\author[Mainz]{Volodymyr Biloshytskyi}
\author[Mainz,HIM]{Xiu-Lei Ren}
\author[Mainz]{Marc Vanderhaeghen}
\address[Mainz]{Institut f\"ur Kernphysik \& PRISMA$^+$  Cluster of Excellence, Johannes Gutenberg Universit\"at,  D-55099 Mainz, Germany}
\address[HIM]{Helmholtz Institut Mainz, D-55099 Mainz, Germany}

\begin{abstract}
In this paper, we present an  improved parameterization of the elastic scattering of spin-0 particles, which is based on a dispersive representation for the inverse scattering amplitude. Besides being based on well known general principles, the requirement that the inverse amplitude should satisfy the dispersion relation significantly constrains its possible forms and have not been incorporated in the existing parameterizations so far. While the right-hand cut of the inverse scattering amplitude is controlled by unitarity, the contribution from the left-hand cut, which comes from the crossing symmetry, is commonly ignored or incorporated improperly. The latter is  parameterized using the expansion in a suitably constructed conformal variable, which accounts for its analytic structure. The correct implementations of the Adler zero and threshold factors for angular momentum $J>0$ are discussed in detail as well. The amplitudes are written in a compact analytic form and provide a useful tool to analyze current and future lattice data in the elastic region improving upon the commonly used Breit-Wigner or K-matrix approaches.

\end{abstract}
%
\end{frontmatter}

\section{Introduction}\label{intro}

Over the last decade there is a renewed interest in the hadron spectroscopy, motivated by recent discoveries of unexpected exotic hadron resonances \cite{LHCb:2020bwg, Aaij:2019vzc,*Aaij:2015tga,Adolph:2014rpp}. At the same time, there is a significant progress in lattice QCD studies of excited hadrons \cite{Briceno:2017max,*Shepherd:2016dni}, which has a great potential for determining the properties of the poorly known hadronic states.

The determination of resonance parameters requires the search for poles in the complex plane. This is particularly important when there is an interplay between several inelastic channels or when the pole is very deep in the complex plane. The fundamental S-matrix constraints, such as unitarity, analyticity, and crossing symmetry significantly constrain the possible form of the amplitude both on the real axis and in the complex plane \cite{Pelaez:2015qba,JPAC:2021rxu,Yao:2020bxx}. Unfortunately, in the experimental and lattice analyses, it is a common practice to ignore most of the S-matrix constraints and rely on the different variants of Breit-Wigner or K-matrix parameterizations due to their simplicity \cite{Briceno:2016mjc,Guo:2018zss,Wilson:2019wfr,*Rendon:2020rtw,Gayer:2021xzv,Cheung:2020mql,Prelovsek:2020eiw,Lang:2014yfa}. Both methods often ignore the existence of the left-hand cut and lead to spurious poles in the complex plane. In addition, for a reaction involving Goldstone bosons, the Adler zero constraint is typically ignored or implemented purely phenomenologically (see e.g. parameterizations suggested in \cite{Bugg:2003kj,Yndurain:2007qm,Pelaez:2016tgi}). As a result, the systematic uncertainties of these approaches are large \cite{Caprini:2008fc}.

The most rigorous way of implementing all S-matrix constraints is to write a fixed-$t$ dispersion relation for the invariant amplitude $T(s,t,u)$, which after projection onto partial waves leads to the set of Roy or Roy-Steiner equations \cite{Roy:1971tc}. In this way, the left-hand cut is treated exactly. In practical applications, however, the rigorous implementation of these equations is almost impossible, since it requires experimental knowledge of all partial waves with different isospin in the direct and crossed channels (including the high-energy region). Therefore, the current precision studies of $\pi\pi$ \cite{GarciaMartin:2011cn,*GarciaMartin:2011jx,*Pelaez:2019eqa,Ananthanarayan:2000ht,*Caprini:2005zr,*Leutwyler:2008xd,Colangelo:2001df} and $\pi K$ \cite{Buettiker:2003pp,*DescotesGenon:2006uk, Pelaez:2020uiw,*Pelaez:2020gnd} scattering are based on a finite truncation, which, in turn, limits the results to the given kinematic region. Furthermore, applying Roy-like equations for coupled-channel cases is relatively complicated and has not been achieved in the literature so far. In \cite{Danilkin:2020pak,*Deineka:2022izb} we used a complementary approach, which is based on solving the partial-wave (p.w.) dispersion relations. In this method, different partial waves with different isospins are fully decoupled, at the expense of the crossing symmetry constraint not being incorporated exactly. The benefit of this approach is that it can be applied to any hadronic reaction, for which the data exist (experimental or lattice), and can be straightforwardly extended to the coupled-channel systems. For instance, one of the central results of \cite{Danilkin:2020pak,*Deineka:2022izb} is the isoscalar coupled-channel $\{\pi\pi,\,K\bar{K}\}$ Omn\`es matrix. It does not have left-hand cuts and therefore serves as the crucial input for the vast production/decay reactions involving the same final states (see e.g. \cite{Danilkin:2018qfn,*Deineka:2019bey,*Danilkin:2019opj,*Danilkin:2021icn} and \cite{Danilkin:2020kce}).

The solution of the p.w. dispersion relation is not available in an analytic form and requires an extra computational cost (especially using the bootstrap technique to estimate the errors),
thus complicating its implementation in fits to lattice data.
However, since some of the resonances are connected almost exclusively to the elastic channels (especially in the lattice calculations with unphysically-large masses for light quarks), it is not always necessary to solve the dispersive integral equation for the direct amplitude. In this work we exploit the idea of writing a dispersion relation for the inverse amplitude and parameterizing the left-hand cut contribution in the physical region by the conformal mapping expansion, which respects its analytical structure. The possible Adler zero of the amplitude is included as an additive pole term in the inverse amplitude.  In the derivation of the dispersive parameterization for the higher partial waves, we implement the angular momentum barrier factor as a relevant kinematic constraint. Even though the proposed parametrizations may look similar to the K-matrix with a Chew-Mandelstam phase space \cite{Wilson:2014cna,Briceno:2016mjc}, or the conformal map parametrizations from \cite{Pelaez:2004vs,Caprini:2008fc,Pelaez:2016tgi}, we will show that it goes beyond those. The requirement that the inverse amplitude satisfies the dispersion relation significantly constrains the possible forms of the parameterization.
Moreover, we discuss in detail the equivalence of the p.w. dispersion relations for the direct and inverse amplitudes for elastic scattering. We also show that the commonly used modified Inverse Amplitude Method (mIAM) \cite{GomezNicola:2007qj,Hanhart:2008mx,Pelaez:2010fj,Nebreda:2010wv}, which satisfies the p.w. dispersion relation for the inverse amplitude, is a special case of the proposed parametrizations.

The paper is organized as follows. We describe the formalism of the dispersive parameterization of the inverse amplitude in Sec.~\ref{sec:formalism}, deriving master formulae for the most relevant cases of elastic scattering. In Sec.~\ref{sec:comparison} we conceptually compare our approach to the commonly used existing methods. Finally, in Sec.~\ref{sec:Numerical examples} we show some numerical test cases, with the emphasis on the recent lattice data on the S-wave isoscalar $\pi\pi\to \pi\pi$ scattering ($m_\pi=239$ MeV\footnote{The estimate of the pion mass on this lattice was improved in \cite{Wilson:2019wfr} resulting in a slight change from the initially reported value of 236 MeV to 239 MeV.}) from the HadSpec collaboration \cite{Briceno:2016mjc}.

\section{Formalism}
\label{sec:formalism}
\subsection{Unitarity}
The $s$-channel partial-wave decomposition for $2\to 2$ process is given by
\begin{equation}\label{p.w.expansion}
T(s,t)=16\pi\,N\,\sum_{J=0}^{\infty}(2J+1)\,t_{J}(s)\,P_J(\cos\theta)\,,
\end{equation}
where $\theta$ is the center-of-mass (c.m.) scattering angle. For the scattering of identical particles $N=2$, while $N=1$ otherwise. This factor is useful to ensure the same unitarity condition for the identical and non-identical two-particle scattering, which in the elastic approximation is given by
\begin{align}\label{Eq:Unitarity}
&\text{Im}\,t_J(s)=\rho(s)\,|t_J(s)|^2\,\theta(s-s_{th})\,, \nonumber \\
&\text{Im}\,\left[t_J(s)\right]^{-1}=-\rho(s)\,\theta(s-s_{th})\,,
\end{align}
where  $s_{th}=(m_1+m_2)^2$ is the threshold. The phase space factor $\rho(s)$ is given by 
\begin{align}\label{Eq:rho}
\rho(s)&=\frac{2\,p(s)}{\sqrt{s}}\,,
\end{align}
where $p(s)$ is the c.m. three-momentum of the system
\begin{align}
p(s)=\frac{\lambda^{1/2}(s,m_1^2,m_2^2)}{2\sqrt{s}}\equiv \sqrt{\frac{(s-m_-^2)(s-m_+^2)}{4s}}\,,
\end{align}
with $\lambda$ being the K\"{a}ll\'{e}n function and $m_{\pm}=m_{1}\pm m_{2}$. At low energy, the effective range expansion for partial waves is conventionally defined as
\begin{align}\label{ERE}
\frac{2}{\sqrt{s_{th}}}\,\text{Re}\,t_J(s) \simeq p^{2J}(s)\,\left(a_J+b_J\, p^2(s)+...\right)\,,
\end{align}
where $a_J$ and $b_J$ denote the scattering length and the slope parameter, respectively.
At high energy, the unitarity condition (\ref{Eq:Unitarity}) guarantees that the partial-wave amplitudes at infinity approach at most constants
\begin{align}
-\frac{1}{2\,\rho(s)}\leq \text{Re}\,t_J(s)\,\leq \frac{1}{2\, \rho(s)},\quad 0\leq \text{Im}\,t_J(s)\leq \frac{1}{\rho(s)}
\end{align}
and in accordance with that, we assume throughout this work that
\begin{align}\label{Eq:tJ_s_to_infty}
t_J(s\to \pm \infty) \to \text{const}\,.
\end{align}

\subsection{S-wave scattering}
\label{SubSec:S-wave}
We start with an investigation of the S-wave ($J=0$) scattering of spinless particles, since in this case we do not need to worry about angular momentum barrier factors. In view of the maximal analyticity assumption \cite{Mandelstam:1958xc,*Mandelstam:1959bc}, the partial-wave amplitude satisfies, according to Eq.~(\ref{Eq:tJ_s_to_infty}), the once-subtracted dispersion relation,
\begin{equation}\label{DRforT_1}
t_0(s)=t_0(s_M)+\frac{s-s_M}{\pi} \int_{L,R}\frac{d s'}{s'-s_M}\frac{\text{Im}\, t_0(s')}{s'-s}+\frac{s-s_M}{s_B-s_M}\frac{g_B^2}{s_B-s}\,.\end{equation}
The symbols $L$ and $R$ denote integrals over left- and right-hand cuts, respectively. The choice of the subtraction point at $s=s_M$ in general is irrelevant and will be discussed later. For completeness, we admitted in Eq.(\ref{DRforT_1}) a possible bound state at $s=s_B$ with the coupling $g_B$, which we will drop later on for simplicity. Using the unitarity relation (\ref{Eq:Unitarity}) on the right-hand cut, Eq.~(\ref{DRforT_1}) simplifies to 
\begin{equation}\label{DRforT_2}
t_0(s) = U(s)+ \frac{s-s_M}{\pi} \int_{s_{th}}^{\infty}\frac{d s'}{s'-s_M}\frac{\rho(s')\,|t_0(s')|^2}{s'-s}\,,
\end{equation}
where we combined the subtraction constant together with the left-hand cut contributions into the function $U(s)$,
\begin{align}\label{Eq:U(s)}
U(s) \equiv t_0(s_M)+\frac{s-s_M}{\pi} \int_{L}\frac{d s'}{s'-s_M}\frac{\text{Im}\, t_0(s')}{s'-s}\,.
\end{align}
The solution to (\ref{DRforT_2}) can be written using the $N/D$ ansatz \cite{Chew:1960iv} 
\begin{align}\label{N/D_1}
t_0(s)=\frac{N(s)}{D(s)}\,,
\end{align}
where the contributions of left- and right-hand cuts are separated into $N(s)$ and $D(s)$ functions, respectively. This ansatz implies a set of linear integral equations \cite{Luming:1964, Johnson:1979jy}. The possible zero of the amplitude, $t_0(s_A)=0$ (which in the case of the $\pi\pi$ and $\pi K$ scattering corresponds to the Adler zero), can be incorporated either as a zero of $N(s)$ or as a pole in $D(s)$. In the former case, the easiest is to choose $s_M=s_A$, leading to
\begin{align}\label{N/D_2}
D(s)&=1- \frac{s-s_A}{\pi} \int_{s_{th}}^{\infty}\frac{d s'}{s'-s_A}\frac{N(s')\,\rho(s')}{s'-s}\,,\\
N(s)&=U(s)+\frac{s-s_A}{\pi} \int_{s_{th}}^{\infty}\frac{d s'}{s'-s_A}\frac{N(s')\,\rho(s')\,(U(s')-U(s))}{s'-s}\,. \nonumber
\end{align}
where $U(s_A)=0$ from (\ref{Eq:U(s)}).
On another side, the set of integral equations with the so-called Castillejo-Dalitz-Dyson (CDD) pole \cite{Castillejo:1955ed} at $s_A$ has the following form (for any $s_M \neq s_A$) 
\begin{align}\label{N/D_3}
D(s)&=1- \frac{s-s_M}{\pi} \int_{s_{th}}^{\infty}\frac{d s'}{s'-s_M}\frac{N(s')\,\rho(s')}{s'-s}+(s-s_M)\frac{g_1}{s-s_A}\,,\nonumber\\
N(s)&=U(s)+\frac{s-s_M}{\pi} \int_{s_{th}}^{\infty}\frac{d s'}{s'-s_M}\frac{N(s')\,\rho(s')\,(U(s')-U(s))}{s'-s} \nonumber\\
&\quad +(s-s_M)\,\frac{g_1\,(U(s)-U(s_A))}{s-s_A}\,.
\end{align}
Using a toy model, both (\ref{N/D_2}) and (\ref{N/D_3}) have been checked numerically to give the same solution that satisfies the initial p.w. dispersion relation (\ref{DRforT_2}) with the constraint $t_0(s_A)=0$. Even though the realization through the CDD pole looks more complicated, it allows us to obtain a simple analytical formula under the assumption that $U(s)\approx U_0=const$. In this case Eq.(\ref{N/D_3}) reduces to
\begin{align}\label{N/D_4}
[t_0(s)]^{-1}& \approx \frac{1}{U_0}+\frac{s-s_M}{\pi} \int_{s_{th}}^{\infty}\frac{d s'}{s'-s_M}\frac{-\rho(s')}{s'-s}\nonumber\\ & \quad +(s-s_M)\frac{g_1/U_0}{s-s_A}\,.
\end{align}
For the more complicated form of the left-hand cut, one needs to solve an integral equation numerically using the matrix inversion method, which requires significant computation time in the case of bootstrap fits.

Alternative to the conventional p.w. dispersion relation, for the elastic scattering one can write a partial wave dispersion relation for the inverse amplitude
\begin{align}\label{DRfor1/T_1}
\left[t_0(s)\right]^{-1}=&\left[t_0(\tilde{s}_M)\right]^{-1}+\frac{s-\tilde{s}_M}{\pi} \int_{L,R}\frac{d s'}{s'-\tilde{s}_M}\frac{\text{Im}\, \left[t_0(s')\right]^{-1}}{s'-s}\nonumber \\
&+\frac{s-\tilde{s}_M}{s_A-\tilde{s}_M}\frac{g_A}{s-s_A}\,,
\end{align}
where we allowed for the pole contribution at $s=s_A$, which corresponds to the Adler zero in $\pi\pi$ and $\pi K$ scattering. The integral over the right-hand cut can be fixed again from unitarity in Eq.~(\ref{Eq:Unitarity}), leading to the following integral equation
\begin{align}\label{DRfor1/T_2}
\left[t_0(s)\right]^{-1}=&\left[t_0(\tilde{s}_M)\right]^{-1}+\frac{s-\tilde{s}_M}{\pi} \int_{L}\frac{d s'}{s'-\tilde{s}_M}\frac{\text{Im}\, \left[t_0(s')\right]^{-1}}{s'-s} \\&+\frac{s-\tilde{s}_M}{\pi} \int_{s_{th}}^{\infty}\frac{d s'}{s'-\tilde{s}_M}\frac{-\rho(s')}{s'-s}+\frac{s-\tilde{s}_M}{s_A-\tilde{s}_M}\frac{g_A}{s-s_A}\,.\nonumber
\end{align}

We emphasize that Eqs. (\ref{DRfor1/T_2}) and (\ref{DRforT_1}) are equivalent: $t_0(s)$ and $\left[t_0(s)\right]^{-1}$ have the same analytic structure, except for the possible presence of the poles (or zeros) in $t_0(s)$ and $\left[t_0(s)\right]^{-1}$. It could be a pole in $t_0(s)$ that correspond to the bound state and therefore $\left[t_0(s_B)\right]^{-1}=0$. Another possibility is a pole in $\left[t_0(s)\right]^{-1}$, that correspond to the Adler zero, $t_0(s_A)=0$. The constraints of $t_0(s_A)=0$ and $\left[t_0(s_B)\right]^{-1}=0$ can be easily incorporated in Eqs. (\ref{DRforT_1}) and (\ref{DRfor1/T_2}) by choosing $s_M=s_A$ and $\tilde{s}_M=s_B$, respectively. We checked numerically for the elastic isoscalar $\pi\pi$ S-wave scattering that Eqs. (\ref{DRforT_1}) and (\ref{DRfor1/T_2}) are consistent with each other using a toy model for the left-hand cut discontinuity in Eq. (\ref{DRforT_1}).

In a general scattering problem, little is known about the left-hand cuts, except their analytic structure in the complex plane. The progress has been made in \cite{Gasparyan:2010xz,*Danilkin:2010xd,*Danilkin:2011fz,*Gasparyan:2012km}. It relies on the consideration of an analytic continuation of the left-hand cut contributions to the physical region, employing a series expansion in terms of a suitably constructed conformal mapping variable $\omega(s)$. The latter is chosen such that it maps the left-hand cut plane onto the unit circle \cite{Frazer:1961zz}. For the most typical cases the exact forms of $\omega(s)$ will be given in Sec.\ref{subsec:conformal_mapping}. In \cite{Danilkin:2020pak,*Deineka:2022izb} the function $U(s)$ of Eq.~(\ref{DRforT_2}) was expanded in the conformal mapping series. Here we suggest to apply the conformal mapping expansion to the first two terms of Eq.~(\ref{DRfor1/T_2}), which leads to the following parametrization of the inverse of $t_0(s)$,
\begin{align}\label{DRfor1/T_finalJ0}
\left[t_0(s)\right]^{-1} \simeq  \sum_{n=0}^{\infty} C_{n}\,\omega^n(s)+R(s,\tilde{s}_M)+\frac{s-\tilde{s}_M}{s_A-\tilde{s}_M}\frac{g_A}{s-s_A}\,,
\end{align}
where the analytical expression of
\begin{align}\label{R(s,s_M)}
R(s,\tilde{s}_M)\equiv & \frac{s-\tilde{s}_{M}}{\pi} \int_{s_{th}}^{\infty}\frac{d s'}{s'-\tilde{s}_{M}}\frac{-\rho(s')}{s'-s}\,,
\end{align}
is given in \ref{Appendix:R}. The unknown coefficients $C_n$, $g_A$ and $s_A$ can be adjusted to reproduce the experimental or lattice data or fixed from the effective field theory by imposing some matching condition. Note that when the conformal series is truncated to a single dominant term, Eq.(\ref{DRfor1/T_finalJ0}) coincides with Eq.(\ref{N/D_4}), which was derived from the $N/D$ ansatz.

The advantage of (\ref{DRfor1/T_2}) compared to (\ref{DRforT_2}) in the elastic approximation is twofold. First of all, when in both dispersion representations the left-hand cut is approximated by the conformal expansion, Eq.~(\ref{DRfor1/T_2}) becomes much simpler than Eq.~(\ref{DRforT_2}), because one does not need to solve numerically the integral equation. Secondly, as it will be shown below, it is easy to extend the formalism to $J\neq 0$. However, the dispersion relation for the inverse amplitude has a clear limitation: it cannot be extended to the coupled-channel case \cite{Badalian:1981xj,GomezNicola:2001as}. Due to the matrix inversion, there would be a mixture of the left-hand cuts of all involved channels, which can also affect the physical region where one has an overlapping cut structure. The typical example is the coupled-channel $\{\pi\pi,K\bar{K}\}$ scattering \cite{Iagolnitzer:1973fq}, in which the left-hand cut of $K\bar{K} \to K\bar{K}$ starts at $4\,(m_K^2-m_\pi^2)$ and the matrix inversion will therefore produce a spurious contribution to the $\pi\pi \to \pi\pi$ unitarity cut. In contrast, the overlapping of the left and right-hand cuts in $K\bar{K} \to K\bar{K}$ does not jeopardize the $N/D$ framework, since it happens in the non-physical region (see Ref. \cite{Danilkin:2020pak,*Deineka:2022izb} for more details and its application to $f_0(980)$).

\subsection{$J \neq 0$, $m_1=m_2$ scattering}
For the case of scattering with $J \neq 0$ one needs to take into account the angular momentum barrier factor 
which implies that around the threshold the amplitude should behave as
\begin{equation}\label{ThresholdConstraintJ=1}
t_J(s)\sim p(s)^{2J} ~  \overset{m_1=m_2} {\sim}  (s-s_{th})^J\,.
\end{equation}
This is implemented by writing a $J+1$ subtracted dispersion relation for the ratio
\begin{equation}
f_J(s)\equiv  \frac{ (s-s_{th})^J}{t_J(s)}\,,
\end{equation}
which is free from kinematic constraints. It leads to
\begin{align}
f_J(s)&=\sum_{i=0}^{J}\frac{1}{i!}\,f_J^{(i)}(s_{th})\,(s-s_{th})^i\\
&+\frac{(s-s_{th})^{J+1}}{\pi}\int_{L,R}\frac{d s'}{(s'-s_{th})^{J+1}}\frac{\text{Im } f_J(s')}{s'-s},\nonumber
\end{align}
where no Adler-related pole was added since there is no known reaction where the amplitude has an extra zero in addition to the one at $s=s_{th}$ given in Eq.~(\ref{ThresholdConstraintJ=1}). Note, that here we subtracted the dispersion relation at the threshold. It allows us to bring the integral over the right-hand cut into the form of Eq.~(\ref{R(s,s_M)}). Indeed, re-expressing $f_J(s)$ in terms of $t_J(s)$ and applying the unitarity relation (\ref{Eq:Unitarity}), we obtain
\begin{align}\label{DRfor1/T_J1_0}
\left[t_J(s)\right]^{-1}=&\frac{1}{(s-s_{th})^J}\sum_{i=0}^{J}\frac{1}{i!}\,f_J^{(i)}(s_{th})\,(s-s_{th})^i\\
&+\frac{s-s_{th}}{\pi} \int_{L}\frac{d s'}{s'-s_{th}}\frac{\text{Im}\, \left[t_J(s')\right]^{-1}}{s'-s}\nonumber \\
&+\frac{s-s_{th}}{\pi} \int_{s_{th}}^{\infty}\frac{d s'}{s'-s_{th}}\frac{-\rho(s')}{s'-s}.\nonumber
\end{align}
Since $f_J^{(i)}(s_{th})$ are in general unknown constants, one can write a general parametrization
\begin{align}
\left[t_J(s)\right]^{-1}\simeq&\frac{1}{(s-s_{th})^J}\sum_{i=0}^{J-1}\frac{1}{i!}\,f_J^{(i)}(s_{th})\,(s-s_{th})^i\\
&+\sum_{n=0}^\infty C_{n}\,\omega^n(s)+R(s,s_{th})\,,\nonumber
\end{align}
where as before, the contribution from the left-hand cut together with the constant term was expanded in a suitably constructed conformal mapping series. For instance, for $J=1$ it corresponds to
\begin{align}\label{DRfor1/T_finalJ1_m1=m2}
\left[t_1(s)\right]^{-1}&\simeq\frac{a}{s-s_{th}}+
\sum_{n=0}^\infty C_{n}\,\omega^n(s)+R(s,s_{th})\,,
\end{align}
and similar for $J=2$
\begin{align}\label{DRfor1/T_finalJ2_m1=m2}
\left[t_2(s)\right]^{-1}&\simeq\frac{a}{(s-s_{th})^2}+\frac{b}{s-s_{th}}+
\sum_{n=0}^\infty C_{n}\,\omega^n(s)+R(s,s_{th}).
\end{align}
The parameters $a$, $b$ and $C_n$ will be fitted to the data in Sec. \ref{sec:Numerical examples}.

\begin{figure}[t]
\centering
\includegraphics[width =0.45\textwidth]{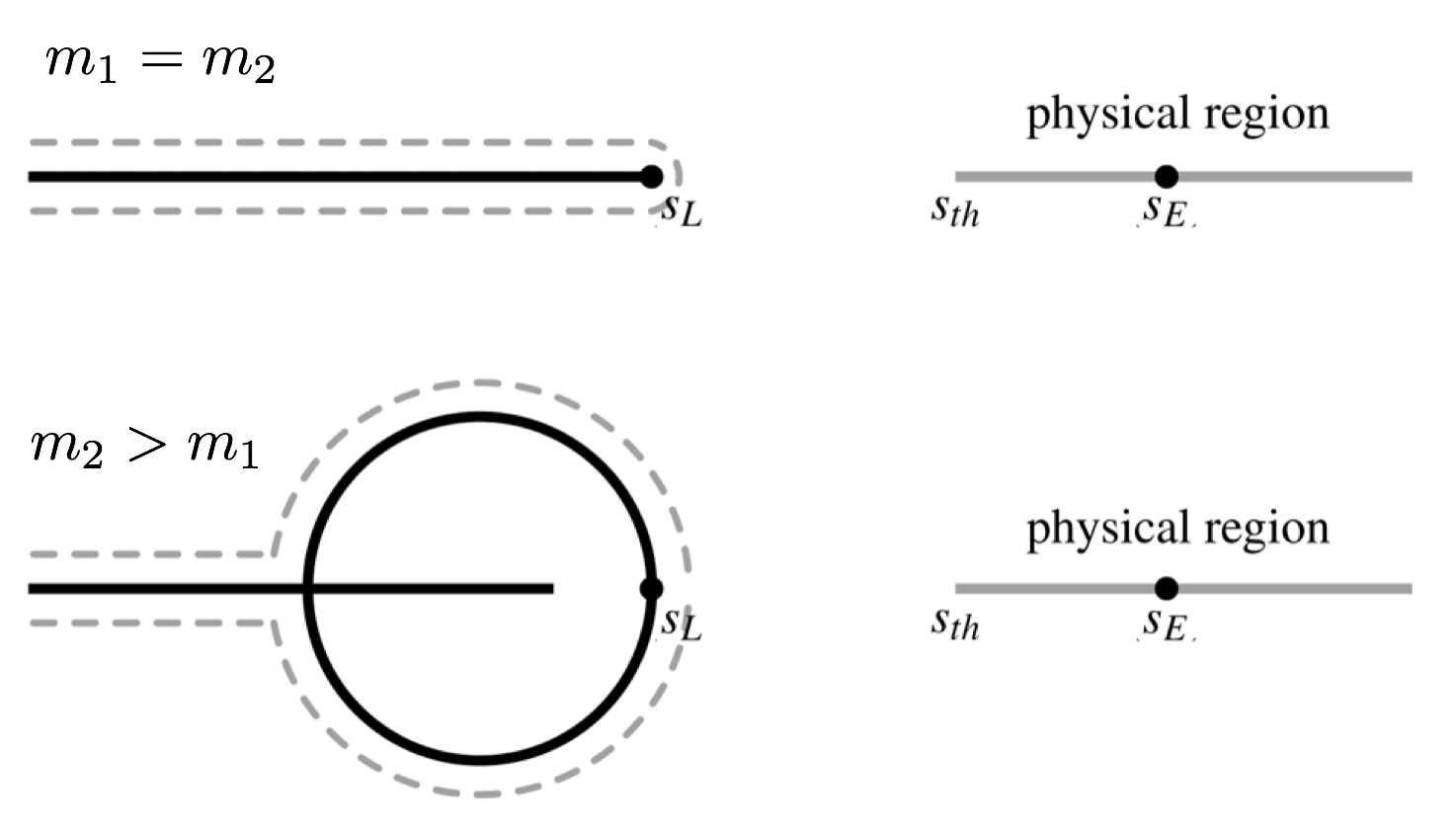}
\caption{Left-hand cut singularities (solid black curves) in the complex $s$-plane for the case when $m_1=m_2$ (upper plane) and $m_2 > m_1$ (lower plane). In the plot, we schematically show the position of the closest left-hand cut singularity ($s_L$), threshold ($s_{th}$), and the expansion point ($s_E$). Dashed lines determine the specific form of the conformal map and subsequently the domain of convergence of the conformal expansion.}
\label{Fig:LeftHandCutStructure}
\end{figure}

\subsection{$J \neq 0$, $m_1 \neq m_2$ scattering}
In case of non-equal masses we propose to write a dispersive representation for the amplitude
\begin{equation}\label{g_J}
g_J(s)\equiv \frac{p(s)^{2J}}{t_J(s)}\,,
\end{equation}
which is free from any kinematic constraints. For simplicity, we limit ourselves here to the $J=1$ case (with an example in mind of $\pi K$ or $\pi D$ scattering), leaving the general result to \ref{Appendix:B}. For $J=1$ we write a twice-subtracted dispersion relation for $g_{1}(s)$ in the following form 
\begin{align}
g_1(s)&=a+b\,s\nonumber\\
&+\frac{(s-m^2_{-})(s-m^2_{+})}{\pi}\int_{L,R}\frac{d s'}{(s'-m^2_{-})(s'-m^2_{+})}\frac{\text{Im}\,g_1(s')}{s'-s},\nonumber
\end{align}
where $m_{\pm}=m_1 \pm m_2$.
Re-expressing $g_1(s)$ in terms of $t_1(s)$ we obtain
\begin{align}\label{DRfor1/T_finalJ1_m1!=m2}
\left[t_1(s)\right]^{-1}&=\frac{a+b\,s}{p^2(s)}+\frac{s}{\pi}\int_{L}\frac{d s'}{s'}\frac{\text{Im}\, \left[t_1(s')\right]^{-1}}{s'-s}\nonumber\\
&\quad +\frac{s}{\pi}\int_{s_{th}}^{\infty}\frac{d s'}{s'}\frac{-\rho(s')}{s'-s}\nonumber\\
&
\simeq \frac{a+b\,s}{p^2(s)}+\sum_{n=1}^\infty C_{n}\,(\omega^n(s)-\omega^n(0))+R(s,0)\,.
\end{align}
Note that for $m_1=m_2$, Eq.~(\ref{DRfor1/T_finalJ1_m1!=m2}) reduces to Eq.~(\ref{DRfor1/T_finalJ1_m1=m2}) with the proper redefinition of the unknown parameters. The analytical expression for $R(s,0)$ is given in Eq.~(\ref{R(s,0)}).

\subsection{Conformal mapping variables $\omega(s)$}
\label{subsec:conformal_mapping}

The form of $\omega(s)$ depends on the cut structure of the reaction. Since it is impossible to write a dispersive representation for the inverse amplitude in the coupled-channel case, we do not include in $\omega(s)$ possible inelastic cuts, as it was proposed in \cite{Pelaez:2004vs,Caprini:2008fc,Pelaez:2016tgi}. Any parameterization for the inverse amplitude which includes an inelasticity cannot be justified from the dispersion relation. As we already pointed out at the end of Sec.\ref{SubSec:S-wave}, the correct way of implementing inelastic cuts is through the coupled-channel dispersion relation for the direct amplitude \cite{Danilkin:2020pak,*Deineka:2022izb} or by considering Roy-like equations \cite{GarciaMartin:2011cn,*GarciaMartin:2011jx,*Pelaez:2019eqa,Ananthanarayan:2000ht,*Caprini:2005zr,*Leutwyler:2008xd,Colangelo:2001df,Buettiker:2003pp,*DescotesGenon:2006uk, Pelaez:2020uiw,*Pelaez:2020gnd}. Therefore $\omega(s)$ is solely specified by the position of the closest left-hand cut branching point ($s_L$) and an expansion point ($s_E$), around which the series is expanded, $\omega(s_E)=0$. The latter typically is chosen in the middle between the threshold and the energy of the last data point that is fitted to the data,
\begin{equation}\label{Eq:s_E}
\sqrt{s_E}=\frac{1}{2}\,\left(\sqrt{s_{th}}+\sqrt{s_{max}}\right)\,.
\end{equation}
This particular choice guarantees a fast convergence of
the conformal expansion in that region. To access some of the systematic uncertainties, the value of $s_E$ can be varied around its central value (\ref{Eq:s_E}).

Since for the scattering of the particles with $m_1=m_2=m$ the left-hand cut lies on the real axis, $-\infty<s<s_L$, one can use a simple function
\begin{align}\label{xi-1}
\omega(s)&=\frac{\sqrt{s-s_L}-\sqrt{s_E-s_L}}{\sqrt{s-s_L}+\sqrt{s_E-s_L}}\,,\\
s_L&=4\,m^2-t_x\,, \nonumber
\end{align}
where $t_x$ is the lowest threshold in the crossed $t$ or $u$ channels. For instance for $\pi\pi \to \pi\pi$ or $D\bar{D}\to D\bar{D}$ scattering $t_x=4\,m_\pi^2$.

For the case when $m_2>m_1$ (e.g. $\pi K \to \pi K$ or $\pi D \to \pi D$ scattering), the left-hand cut structure is a bit more complicated (see Fig.~\ref{Fig:LeftHandCutStructure}). In addition to the left-hand cut lying on the real axis $-\infty < s < (m_2 - m_1)^2 $, there is a circular cut at $|s|=m_2^2 - m_1^2$. The conformal map that meets these requirements is defined as
\begin{align}\label{xi-2}
\omega(s)&=-\frac{\left(\sqrt{s}-\sqrt{s_E}\right)\left(\sqrt{s} \sqrt{s_E}+s_L\right)}{\left(\sqrt{s}+\sqrt{s_E}\right) \left(\sqrt{s} \sqrt{s_E}-s_L\right)}\,,\\
s_L&=m_2^2 - m_1^2\,. \nonumber
\end{align}
We note that for the forms of $\omega(s)$,  given in Eqs. (\ref{xi-1}) and (\ref{xi-2}), the conformal series, being truncated at any finite order, is bounded asymptotically. This is consistent with the assigned asymptotic behavior.

\section{Comparison to other parametrizations}
\label{sec:comparison}
Before applying dispersively justified representations to the physical cases, it is instructive to compare Eqs.~(\ref{DRfor1/T_finalJ0},\,\ref{DRfor1/T_finalJ1_m1=m2},\,\ref{DRfor1/T_finalJ2_m1=m2},\,\ref{DRfor1/T_finalJ1_m1!=m2}) with the commonly used parametrizations.

\subsection{K-matrix approach}
The K-matrix approach can be written in a general form as 
\begin{equation}\label{Eq:KmatrixGeneral}
\left[t_J(s)\right]^{-1}=\frac{1}{p(s)^{2J}}K^{-1}(s)+I(s)\,,
\end{equation}
where $\text{Im}\,I(s)=-\rho(s)$. The standard K-matrix implementations correspond to \cite{Zyla:2020zbs}
\begin{equation}\label{Eq:Kmatrix1}
K(s)=\frac{g^2}{m^2-s}+\sum_n \gamma_n\,s^n\,,
\end{equation}
with $I(s)$ being the conventional phase-space $I(s)=-i\,\rho(s)$ or its Chew-Mandelstam \cite{Chew:1960iv} version, $I(s)=I(s_{M})+R(s,s_{M})$. The possible Adler zero is typically added by weighting $K(s)$ by a factor $(s-s_A)$ \cite{Briceno:2016mjc},
\begin{equation}\label{Eq:Kmatrix2}
K(s)=(s-s_A)\,\left(\frac{g^2}{m^2-s}+\sum_n \gamma_n\,s^n\,\right).
\end{equation}
The attempt of adding the left-hand cut contribution was made in \cite{Yndurain:2007qm}, using the conformal mapping expansion. The suggested parameterization for the case with Adler zero was written as
\begin{equation}\label{Eq:Pelaez1}
K^{-1}(s)=\frac{m_\pi^2}{s-s_A}\left(
\frac{2s_A}{m_\pi\sqrt{s}}+\sum_{n=0}^\infty C_{n}\,\omega^n(s)\right)
\end{equation}
or in an alternative form as \cite{Caprini:2008fc,Pelaez:2016tgi}
\begin{equation}\label{Eq:Caprini}
K^{-1}(s)=\frac{m_\pi^2}{s-s_A}\left(\sum_{n=0}^\infty C_{n}\,\omega^n(s)\right)\,.
\end{equation}

First of all, we emphasize that all the parametrizations with conventional phase-space are at best limited to the physical region only. Below threshold, they have unphysical left-hand singularities and cannot be connected to the dispersion relation. The non-analytic behaviour of $\rho(s)$ below the two-particle threshold also often causes spurious poles in the complex plane.
The attempt of fixing this problem by adding an additional $\sqrt{s}$ term in Eq.(\ref{Eq:Pelaez1}) only place this parameterization further away from the dispersive construction. In the parametrizations with Chew-Mandelstam phase-space, the subtraction point $s_M$ and the constant $I(s_{M})$ are typically freely chosen. However, in some cases it can essentially affect the structure of the left-hand cut contribution. For instance, for
$J=1$ and $m_1 \neq m_2$ the particular form of Chew-Mandelstam phase space with $s_M=0$ is connected to the particular form of the left-hand cut contribution (see Eq.(\ref{DRfor1/T_finalJ1_m1!=m2})). 

The commonly used K-matrix implementations given by Eqs. (\ref{Eq:Kmatrix1}, \ref{Eq:Kmatrix2}) at best assume that the left-hand cut contribution can be approximated by a constant. The inclusion of the left-hand cuts in Eqs.\,(\ref{Eq:Pelaez1}, \ref{Eq:Caprini}) is a step forward, however its implementation is non-dispersive for two main reasons. First, the left-hand cut contribution in Eqs.\,(\ref{Eq:Pelaez1}, \ref{Eq:Caprini}) is multiplied by the pole contribution from the Adler zero. The correct implementation of the analytic properties, in turn, requires additive contribution from the left-hand cut and the pole due to Adler zero as derived in Eqs.\,(\ref{DRfor1/T_2},\ref{DRfor1/T_J1_0},\ref{DRfor1/T_finalJ1_m1!=m2}). Second, the proposed form of $\omega(s)$ in \cite{Yndurain:2007qm,Caprini:2008fc,Pelaez:2016tgi} accounts not only for the left-hand cut, but also for the inelastic cuts. This more general form of $\omega(s)$ can only help with an effective description around the inelastic threshold. It does not improve the validity of the parameterization in the complex plane, since both on the left-hand and inelastic cuts the conformal mapping expansion does not converge by construction, i.e. $\omega(s)=1$. As emphasized before, it is not possible to write a dispersion relation for the inverse amplitude in the coupled-channel case and therefore the inclusion of the contribution from the inelastic cuts into $\omega(s)$ does not have a firm dispersive ground.

\subsection{mIAM and IAM}
Finally, let's compare Eqs. (\ref{DRfor1/T_2}) and  (\ref{DRfor1/T_J1_0}) with the IAM (mIAM) \cite{Truong:1988zp,*Dobado:1989qm,*Truong:1991gv,*Dobado:1992ha,GomezNicola:2007qj}. The latter is a widely used way of unitarizing $\chi$PT (see e.g. Refs.\cite{Hanhart:2008mx,Pelaez:2010fj,Nebreda:2010wv,GomezNicola:2001as,Dai:2011bs}). For the elastic S-wave scattering it has the following form
\begin{align}\label{mIAM}
t_0^{\text{mIAM}}(s)=\frac{\left[t_0^{\text{LO}}(s)\right]^2}{t_0^{\text{LO}}(s)-\left[t_0^{\text{NLO}}(s)-t_0^{\text{LO}}(s)\right]+A^{\text{mIAM}}(s)}\,,
\end{align}
where $t_0^{\text{LO}}$ and $t_0^{\text{NLO}}$ are leading order (LO) and next-to-leading order (NLO) S-wave scattering amplitudes in $\chi$PT, respectively. Note, that Eq.~(\ref{mIAM}) can be naively derived in the physical region by performing NLO expansion of $\text{Re} [t(s)]^{-1}$ and plugging it into $[t(s)]^{-1}=\text{Re} [t(s)]^{-1}-i\,\rho(s)$ relation. On the other hand, it has been shown in Ref. \cite{GomezNicola:2007qj} that Eq.~(\ref{mIAM}) can be justified by writing the p.w. dispersion relation for the inverse amplitude and approximating the subtraction constants and the left-hand cut discontinuity by its chiral expansion. The $A^{\text{mIAM}}(s)$ term in needed to remove a spurious pole on the real axis below threshold and at the same time incorporate correctly the Adler zero.

By taking LO and NLO SU(2) $\chi$PT amplitudes from \cite{Niehus:2020gmf} with low energy constants from \cite{Bijnens:2014lea} for the isoscalar S-wave $\pi\pi$ scattering we have checked that mIAM given in Eq.~(\ref{mIAM}) indeed satisfies the dispersion relation for the inverse amplitude (\ref{DRfor1/T_2})
\begin{align}\label{DRfor1/T_mIAM}
&\left[t^{\text{mIAM}}_0(s)\right]^{-1}=
\left[t^{\text{mIAM}}_0(\tilde{s}_{M})\right]^{-1}
+\frac{s-\tilde{s}_{M}}{s_A-\tilde{s}_{M}}\frac{g_A}{s-s_A}\nonumber \\
&\qquad \qquad 
+\frac{s-\tilde{s}_{M}}{\pi} \int_{L,R}\frac{d s'}{s'-\tilde{s}_{M}}\frac{\text{Im}\, \left[t^{\text{mIAM}}_0(s')\right]^{-1}}{s'-s},\nonumber\\
& g_A=\left(\left.\frac{\text{d} t^{\text{mIAM}}_0(s)}{\text{d} s}\right\vert_{s=s_A}\right)^{-1}\,.
\end{align}
This implies that mIAM is a specific case of the proposed parameterization given in Eq.(\ref{DRfor1/T_finalJ0}). We have checked that for fixed $s_A$, $g_A$ to the mIAM and by adjusting just one leading term in the conformal expansion $C_0=t^{\text{mIAM}}_0(\tilde{s}_M=s_{th})$ we can reproduce the results of the mIAM in the physical region and in the complex plane around $\sigma/f_0(500)$ resonance with $\sim 5\%$ accuracy.

\begin{figure}[h]
\centering
\includegraphics[width =0.40\textwidth]{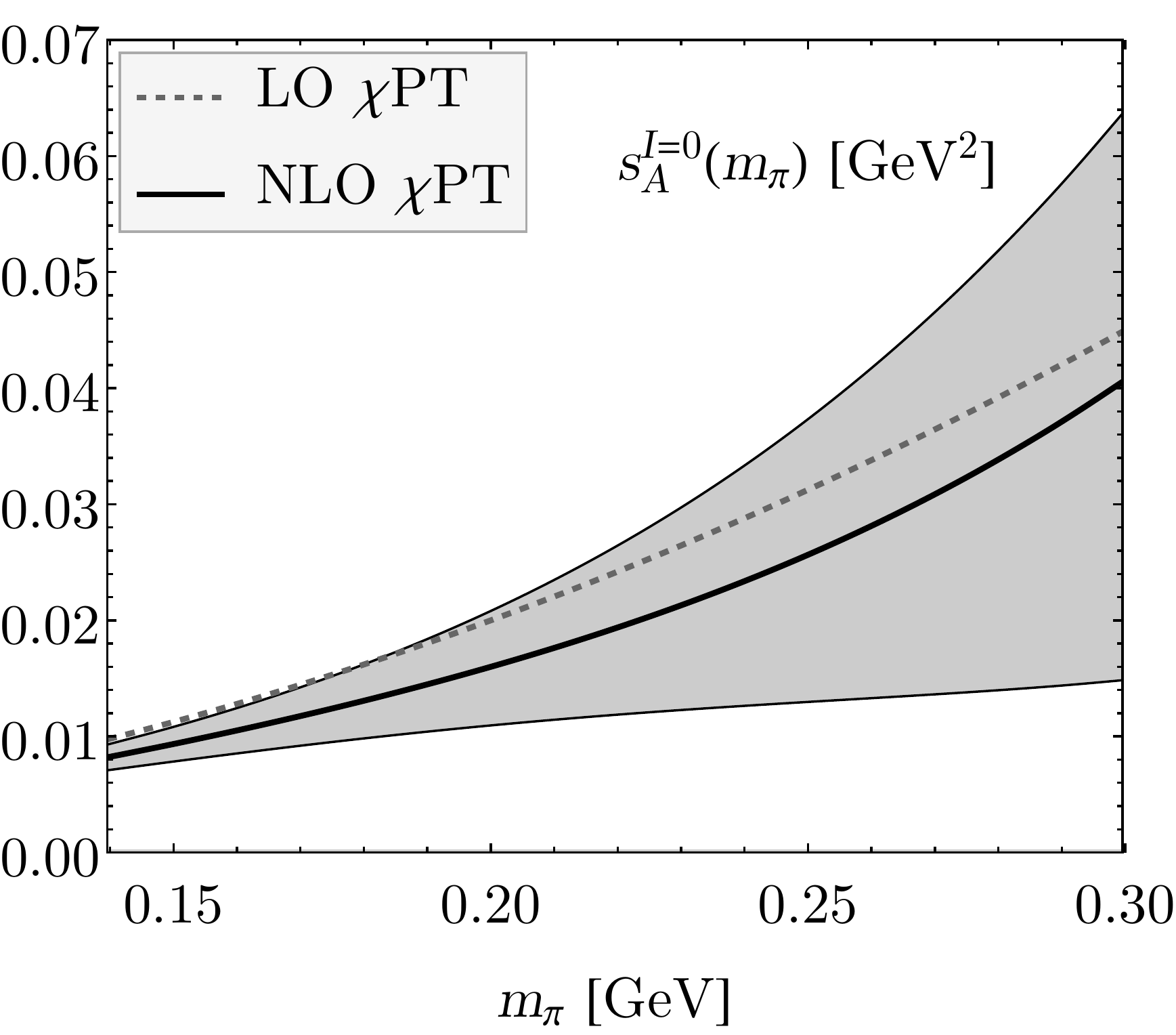}
\caption{Adler zero position $s_A$ as a function of the pion mass for
the S-wave isoscalar $\pi\pi$ scattering. In the plot we used SU(2) $\chi$PT p.w. amplitudes at leading order (LO) and next-to-leading order (NLO) from \cite{Niehus:2020gmf} (expressed in terms of the pion decay constant in the chiral limit) and low-energy constants (LECs) $l_1^r = -4.03(63)\times 10^{-3}$, $l_2^r = 1.87(21)\times 10^{-3}$, $l_3^r = 0.8(3.8)\times 10^{-3}$, $l_4^r = 6.2(1.3)\times 10^{-3}$ from \cite{Bijnens:2014lea}.}
\label{Fig:sA}
\end{figure}

Note that it is not guaranteed by the dispersion construction (see e.g. Eq.(\ref{DRfor1/T_2}) or Eq.(\ref{DRfor1/T_mIAM})) that the inverse amplitude would not turn zero somewhere in the complex plane leading to unphysical poles of the amplitude itself. That is exactly what happens with mIAM, which does not satisfy the dispersion relation for the direct amplitude (\ref{DRforT_2}) due to two spurious poles on the 1st Riemann sheet at $s=-0.87\pm 0.49\,i\,\, \text{GeV}^2$. These spurious poles are an artifact of a specific model for the left-hand cuts. In general, the mIAM, or the more general parameterization given in Eq.(\ref{DRfor1/T_finalJ0}), is not meant to be applied in the unphysical region of the 1st Riemann sheet. What is more important, is that there are no spurious poles on the 2nd Riemann sheet, where the $\sigma/f_0(500)$ pole resides.

For the P-wave $\pi\pi$ scattering, we checked that $t^{\text{IAM}}_1(s)$ given by \cite{Truong:1988zp,*Dobado:1989qm,*Truong:1991gv,*Dobado:1992ha}
\begin{align}
t_1^{\text{IAM}}(s)=\frac{\left[t_1^{\text{LO}}(s)\right]^2}{t_1^{\text{LO}}(s)-\left[t_1^{\text{NLO}}(s)-t_1^{\text{LO}}(s)\right]},
\end{align}
satisfies the p.w. dispersion relations for the inverse amplitude given by Eq.~(\ref{DRfor1/T_J1_0}), i.e. 
\begin{align}\label{DRfor1/T_IAM}
&\left[t^{\text{IAM}}_1(s)\right]^{-1}=\frac{f^{\text{IAM}}_1(s_{th})}{s-s_{th}}+\left[f^{\text{IAM}}_1(s_{th})\right]'\\
&+\frac{s-s_{th}}{\pi} \int_{L,R}\frac{d s'}{s'-s_{th}}\frac{\text{Im}\, \left[t^{\text{IAM}}_1(s')\right]^{-1}}{s'-s}\,,~~ f^{\text{IAM}}_1(s) \equiv \frac{s-s_{th}}{t^{\text{IAM}}_1(s)}\,. \nonumber
\end{align}
Here, similarly to the S-wave, one can adjust the parameters of the conformal expansion in Eq.(\ref{DRfor1/T_finalJ1_m1=m2}) such that Eq.(\ref{DRfor1/T_finalJ1_m1=m2}) reproduces IAM almost exactly. We have also checked that in P-wave IAM does not have any spurious poles.

\begin{table*}
\begin{tabular*}{\textwidth}[t]{@{\extracolsep{\fill}}l|lllll|c|l|l@{}}
        \hline \hline
        & \multicolumn{5}{c|}{Fit parameters}  &  &  \multicolumn{2}{c}{Pole position} \\
        & $g_A$ & $a$ & $b$ & $C_0$ & $C_1$  & $\chi^2/\text{d.o.f}$ & $\sqrt{s_p}$, MeV & $\sqrt{s^{\text{Roy-like}}_p}$, MeV \\
        \hline\hline
\multicolumn{9}{l}{$\pi \pi \to \pi \pi$}\\
\hline
        $(J=0,I=0)$, Exp        & 0.44 & - & - & 4.86 & - & 0.4 & $468-i\,239$& $457^{+14}_{-13}-i\,279^{+11}_{-7}$\\
        \hline
        $(J=0,I=0)$, Lattice  & 0.56 & - & - & 1.37 & - & 0.8 & $560-i\,169$ & - \\
        $m_\pi = 239$ MeV & & & & & & & &\\
\hline
        $(J=0,I=2)$, Exp        & -0.84 & - & - & -23.98 & - & 0.0 & - & - \\  
\hline
        $(J=1,I=1)$, Exp        & - & 2.30 & - & -5.42 & - & 3.0 & $758 - i\,73$& $763.7^{+1.7}_{-1.5} - i\,73.2^{+1.0}_{-1.1}$ \\
                                & - & 1.92 & - & -4.26 & -4.44 & 0.9 & $762 - i\,71$&  \\
\hline
        $(J=2,I=0)$, Exp        & - & 0.04 & 10.01 & -7.87 & - & 1.1 & $1261 - i\,94$& $1267.3^{+0.9}_{-0.9} - i\, 87(9)$\\  
        \hline
\multicolumn{9}{l}{$\pi K \to \pi K$}\\
\hline
        $(J=0,I=1/2)$, Exp      & 0.44 & - & - & 2.30 & - & 2.1 & $707 - i\,246$ & $648(7) - i\,280(16)$\\
                                & 0.22 & - & - & 1.48 & 1.54 & 0.0 & $684- i\,312$ &\\
\hline
        $(J=0,I=1/2)$, Lattice  & 0.63 & - & - & 1.53 & - & 0.4 & $764 - i\,278$ & -\\
        $m_\pi = 239$ MeV & & & & & & & &\\
\hline
        $(J=0,I=3/2)$, Exp      & -0.86 & - & - & -7.86 & - & 0.5 & - & - \\  
\hline
        $(J=1,I=1/2)$, Exp       & - & 0.86 & -1.05 & - & - & 0.7 & $889-i\,27$ & $890(2)-i\,25.6(1.2)$\\
        \hline\hline 
    \end{tabular*}
\caption{Fit parameters entering Eqs.~(\ref{DRfor1/T_finalJ0},\,\ref{DRfor1/T_finalJ1_m1=m2},\,\ref{DRfor1/T_finalJ2_m1=m2},\,\ref{DRfor1/T_finalJ1_m1!=m2}) which were adjusted to reproduce available pseudo-data from the Roy-like analyses \cite{GarciaMartin:2011cn,*GarciaMartin:2011jx,*Pelaez:2019eqa, Pelaez:2020uiw,*Pelaez:2020gnd} or lattice data \cite{Briceno:2016mjc,Wilson:2019wfr}. In Eq.~(\ref{DRfor1/T_finalJ0}), the Adler position $s_A$ is fixed from the LO $\chi$PT (given in Eq.~(\ref{Eq:sAdler_LO})), while the subtraction constant is chosen to be at threshold, $\tilde{s}_M=s_{th}$. In the right columns we collect pole positions found on the 2nd Riemann sheet and compare them with the Roy-like extractions \cite{Pelaez:2015qba,GarciaMartin:2011cn,*GarciaMartin:2011jx,*Pelaez:2019eqa, Pelaez:2020uiw,*Pelaez:2020gnd}.
See text for more details.}
\label{tab:FitResults}
\end{table*}

\begin{figure*}[t]
\centering
\includegraphics[width =0.33\textwidth ]{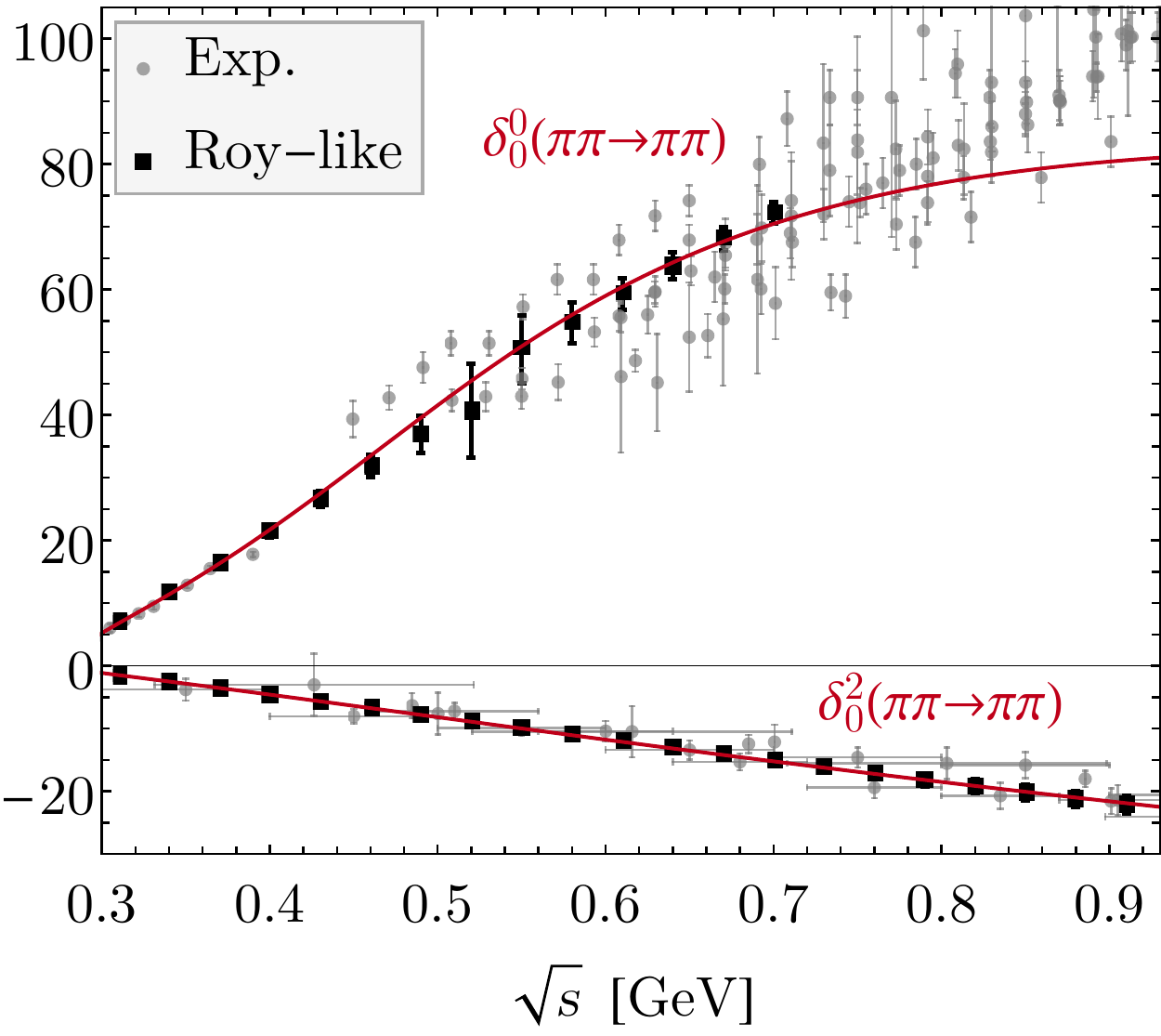}
\includegraphics[width =0.33\textwidth ]{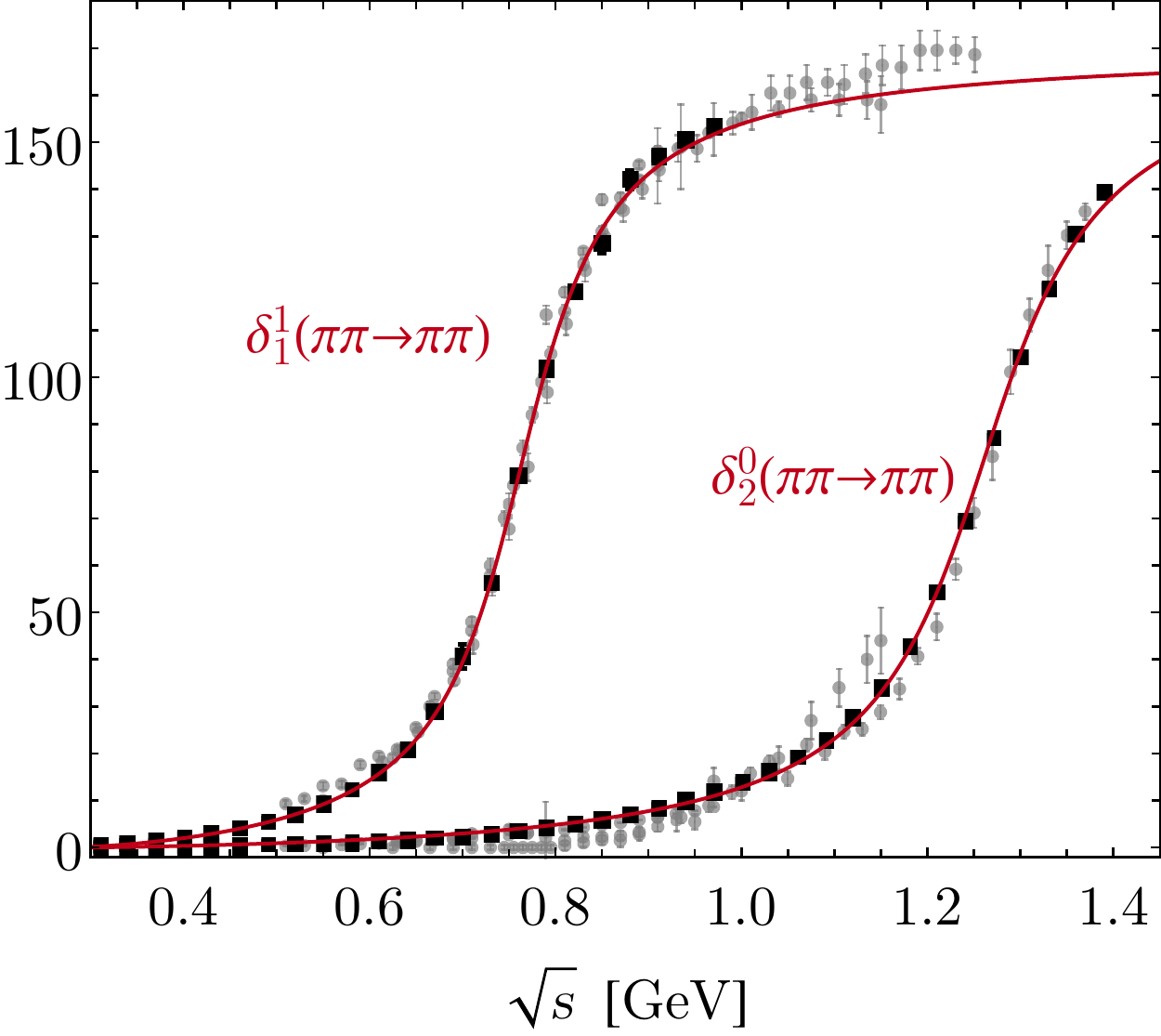}
\includegraphics[width =0.33\textwidth ]{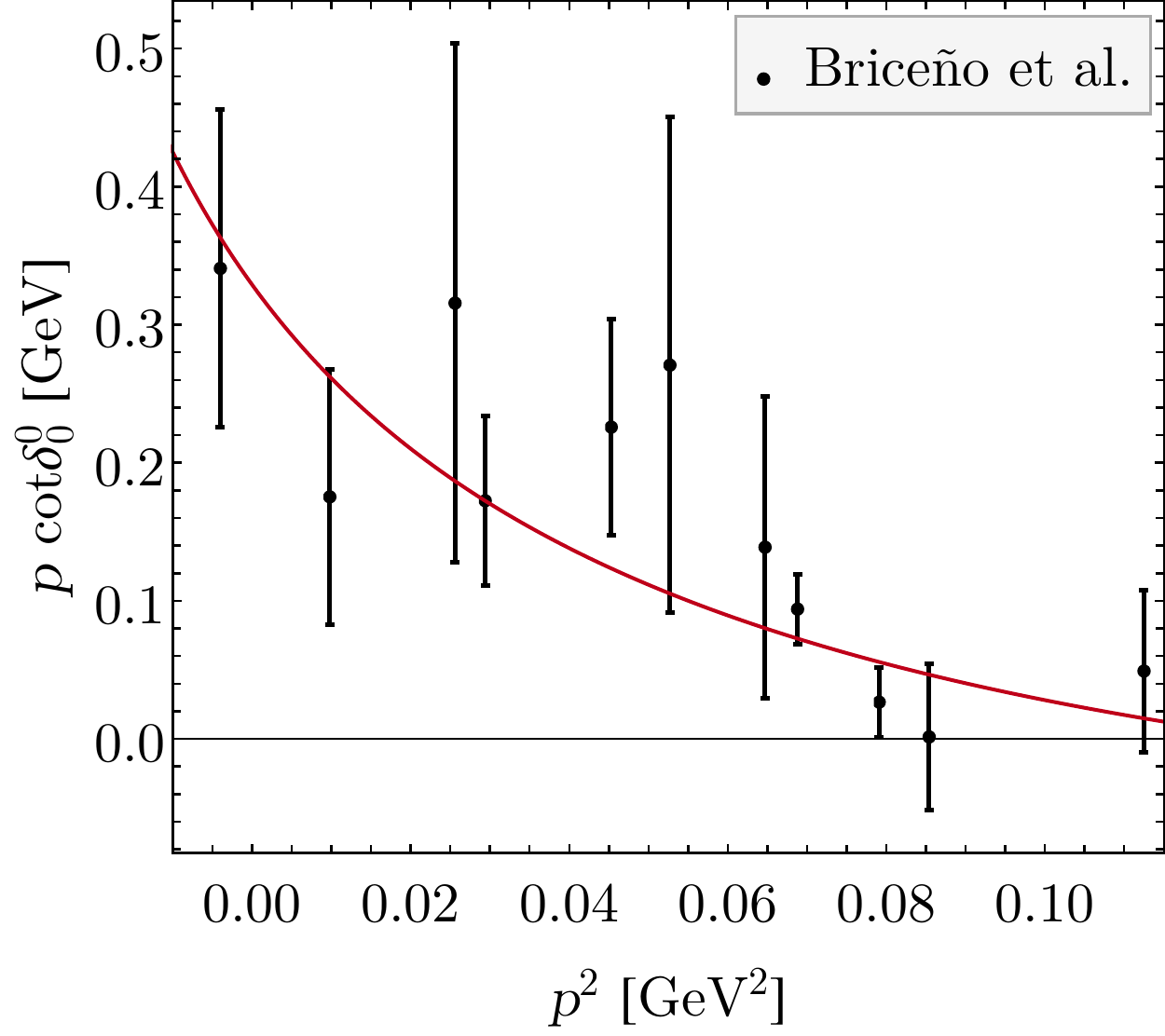}\\
\includegraphics[width =0.33\textwidth ]{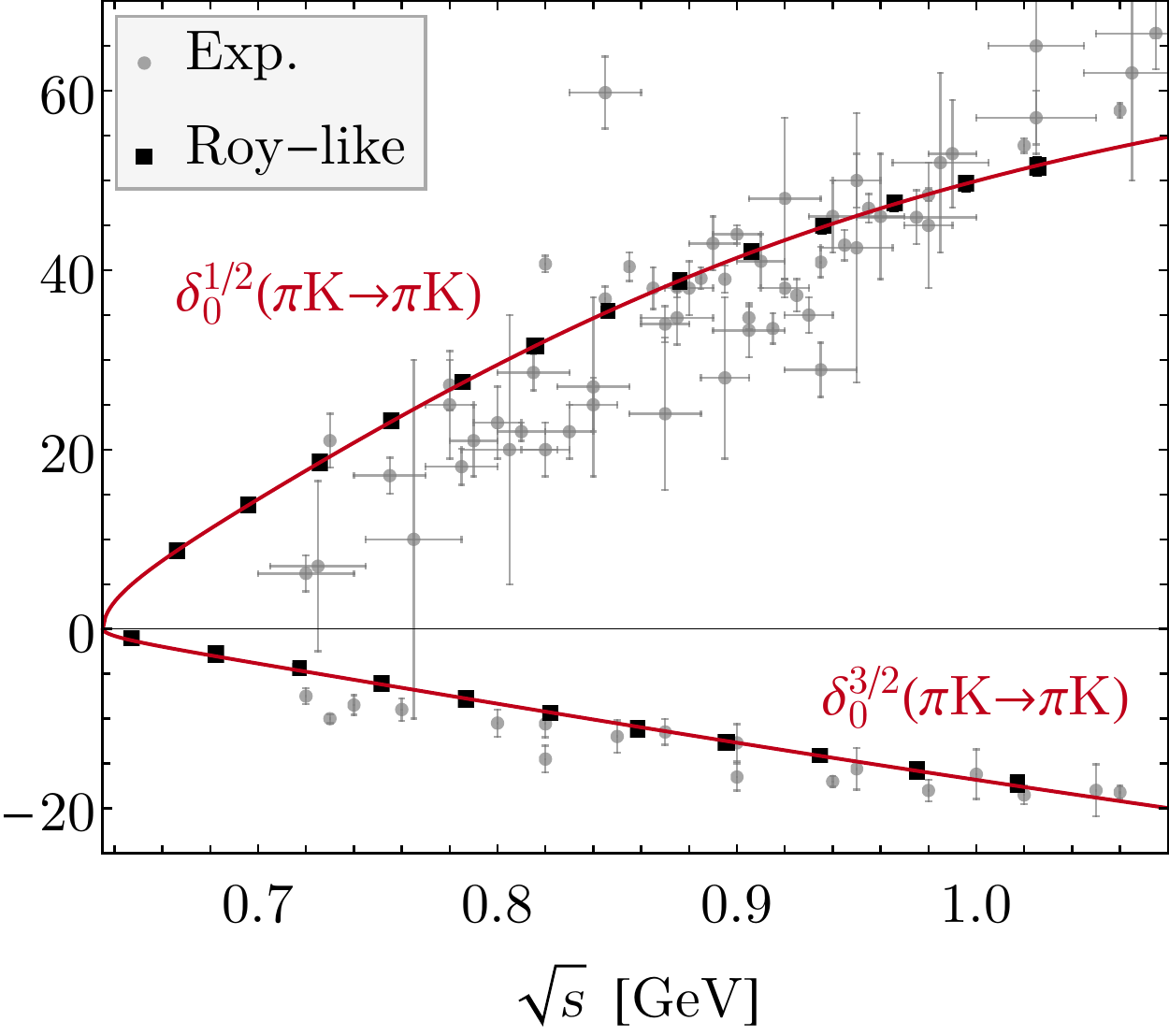}
\includegraphics[width =0.33\textwidth ]{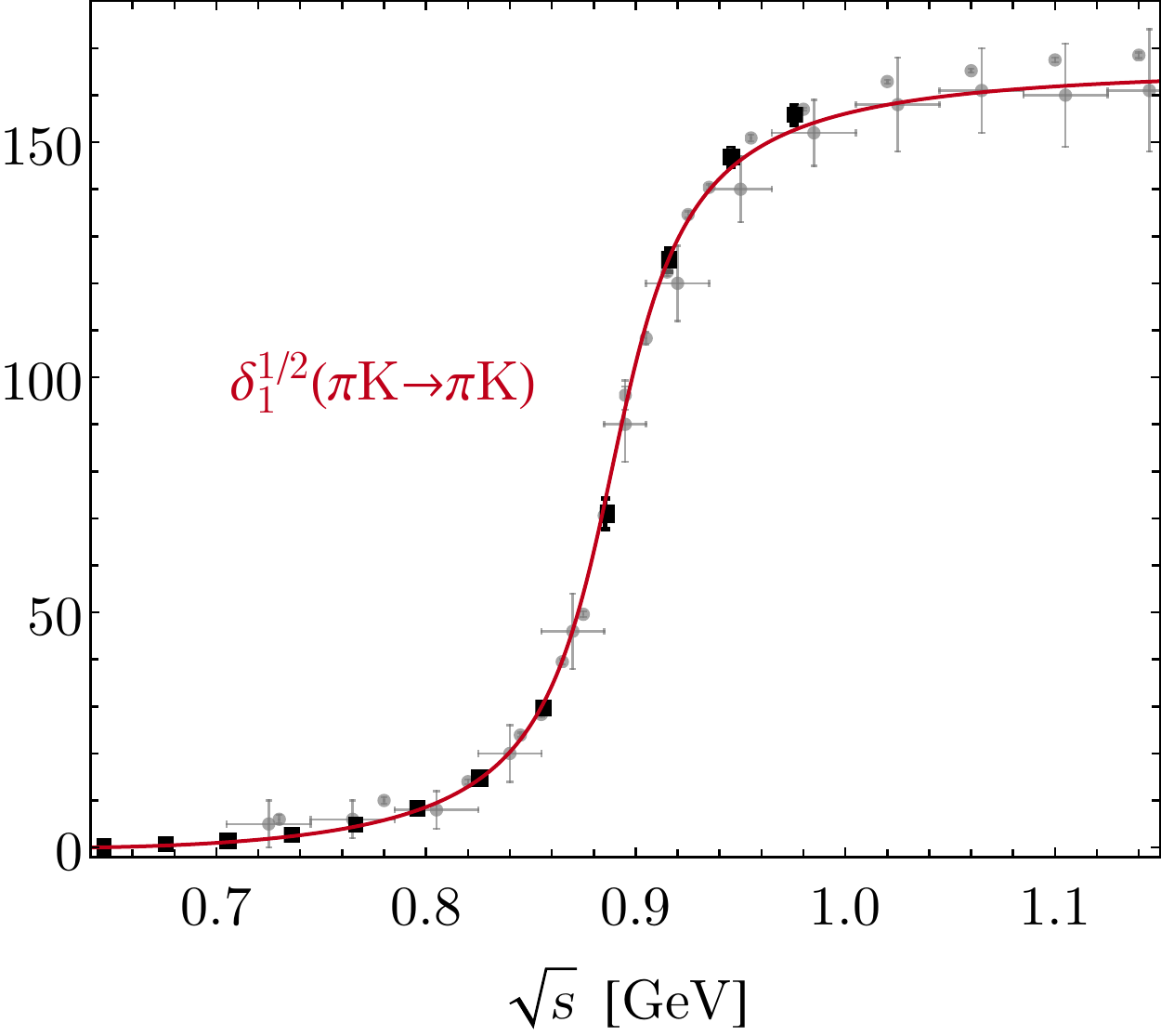}
\includegraphics[width =0.33\textwidth ]{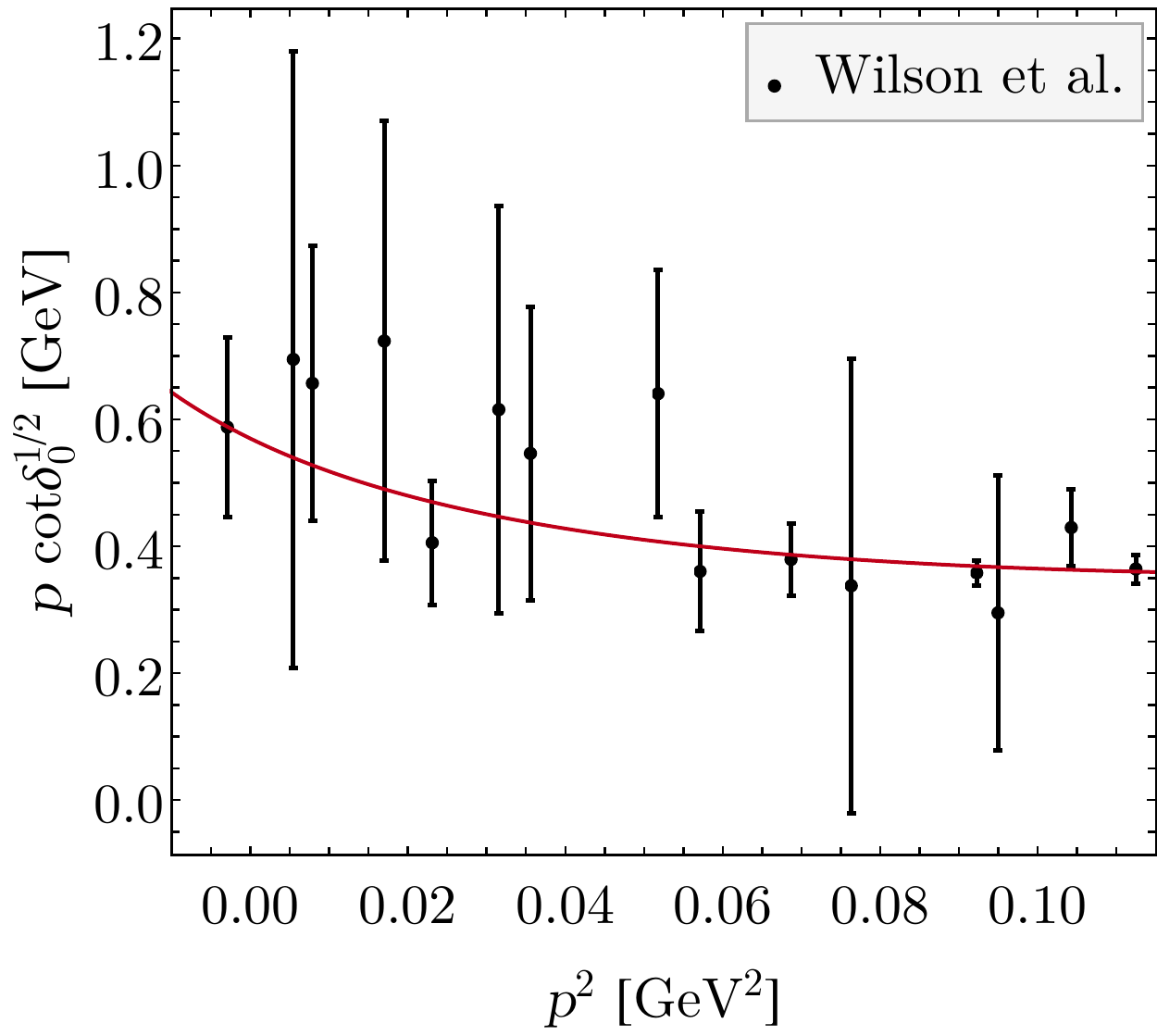}
\caption{Results of the fits ($\delta^I_J$) to the pseudo-data from the Roy-like analyses (left and central panels) \cite{GarciaMartin:2011cn,*GarciaMartin:2011jx,*Pelaez:2019eqa,Pelaez:2020uiw,*Pelaez:2020gnd} and lattice data (right panels) \cite{Briceno:2016mjc,Wilson:2019wfr}. In all cases (except $\delta_0^0$ and $\delta_0^2$ and fits to lattice data) the last fitted data point is around $\sim 1$ GeV, which corresponds to the validity of Roy-like solution. In the case of $\delta_0^0$, however, we limited the fit till $700$ MeV, since above the effect of the $f_0(980)$ becomes important. For $\delta_0^2$ we took $s^{1/2}_{max}=1.42$ GeV, which correspond to the highest point where the set of forward dispersion relations was applied \cite{GarciaMartin:2011cn,*GarciaMartin:2011jx,*Pelaez:2019eqa, Pelaez:2020uiw,*Pelaez:2020gnd}.}
\label{Fig:Results}
\end{figure*}

The main difference between the proposed dispersive inverse amplitudes given in Eqs.~(\ref{DRfor1/T_finalJ0},\,\ref{DRfor1/T_finalJ1_m1=m2},\,\ref{DRfor1/T_finalJ2_m1=m2},\,\ref{DRfor1/T_finalJ1_m1!=m2}) and mIAM (IAM) is that the former can be applied to any elastic scattering, and it is not limited to the Lagrangian based resummation scheme \cite{Pelaez:2021dak}. The unknown parameters in Eqs.~(\ref{DRfor1/T_finalJ0},\,\ref{DRfor1/T_finalJ1_m1=m2},\,\ref{DRfor1/T_finalJ2_m1=m2},\,\ref{DRfor1/T_finalJ1_m1!=m2}) can be fixed directly from the experimental or lattice data (some numerical examples will be shown below). The only required input for the $\pi\pi$ and $\pi K$ scattering is the Adler zero position, since it lies in the unphysical region and typically can not be constrained well by the data. On another side, the unknown parameters in Eqs.~(\ref{DRfor1/T_finalJ0},\,\ref{DRfor1/T_finalJ1_m1=m2},\,\ref{DRfor1/T_finalJ2_m1=m2},\,\ref{DRfor1/T_finalJ1_m1!=m2}) can be estimated from $\chi$PT (an example is shown at the end of Section \ref{sec:Numerical examples}), which allows to connect continuously the results with different quark masses, similar to mIAM.

\section{Adler zero}
For the S-wave $\pi\pi$ and $\pi K$ scattering one has to account for an Adler zero of the amplitude required by chiral symmetry. Its position typically lies very close to the left-hand cut and cannot be determined precisely from the fit to data in the physical region. Around the Adler zero the chiral perturbation theory ($\chi$PT) converges relatively fast. For the physical pion mass the higher-order corrections shifts the 
LO results 
\begin{align}\label{Eq:sAdler_LO}
&s_A^{I=0}=\frac{m_\pi^2}{2},~~ s_A^{I=2}=2\,m_\pi^2,~~s_A^{I=3/2}=m_\pi^2+m_K^2\,,  \nonumber\\
&s_A^{I=1/2}=\frac{1}{5}\left(m_\pi^2+m_K^2+2\sqrt{4\,(m_K^2-m_\pi^2)^2+m_\pi^2\,m_K^2}\right)\,.
\end{align}
only slightly. For larger than physical pion mass values the position of the Adler zero becomes more sensitive to the input from the LECs, which enters at higher orders. As an example, in Fig.~\ref{Fig:sA} we show the isoscalar Adler zero position as a function of the pion mass up to NLO order. The maximal $m_\pi$ is considered to be $\sim 300$ MeV, which lies well within the $\chi$PT range of applicability. The LECs were taken from \cite{Bijnens:2014lea} and assumed to be uncorrelated. This assumption is rather strong and effectively includes an uncertainty related to the truncation error of the chiral expansion. 

Since the lattice data do not provide enough information to control the amplitude around the Adler zero, it is important to implement this chiral constraint in the fits. As it will be shown in the next section, for $m_\pi=239$ MeV the error on the $f_0(500)$ pole parameters due the Adler zero position is suppressed compared to the statistical error from the lattice data \cite{Briceno:2016mjc}. For pion masses sufficiently larger 
than $300$ MeV, the Adler zero constraint may be less accurate. In this case, a reasonable strategy is to perform the fit both with and without the Adler zero constraint.

\section{Numerical examples}
\label{sec:Numerical examples}

In this section, we present some test fits to the well-established $\pi\pi$ and $\pi K$ scattering with $J=0,1,2$ in the low energy region. Here we do not attempt to provide a detailed analysis of experimental data. Instead, we opt for fitting the result of the Roy (Roy-Steiner) analyses \cite{GarciaMartin:2011cn,*GarciaMartin:2011jx,*Pelaez:2019eqa, Pelaez:2020uiw,*Pelaez:2020gnd}, as the best representation of the data. The goal is to show that the proposed dispersive parametrizations are suitable to the search for poles in the complex plane and can describe both: wide tetraquark states (like $\sigma/f_0(500)$) and relatively narrow quark-antiquark states (like the $\rho(770)$ or $f_2(1270)$ mesons). For simplicity, we present our numerical test results using the LO input (\ref{Eq:sAdler_LO})

As one can see in Table \ref{tab:FitResults} and Fig. \ref{Fig:Results}, an accurate description of the Roy (Roy-Steiner) pseudo-data is achieved with at most 3 parameters. We also observed, that adding more terms in the conformal expansion, one can systematically improve the fits. Note that by fitting the Roy (Roy-Steiner) results, which are smooth functions, the $\chi^2/d.o.f$ loses its statistical meaning and can be $< 1$.

As an example, we also perform a numerical comparison between Eq.(\ref{DRfor1/T_finalJ0}) and the most advanced analytical  $\sigma/f_0(500)$ parameterization given by Eqs. (7, 10, 12, 14) of Ref.\,\cite{Caprini:2008fc}. That parameterization corresponds to Eqs.(\ref{Eq:KmatrixGeneral},\ref{Eq:Caprini}) with $I(s)=R(s,0)$ and the conformal variable, 
\begin{equation}\label{Eq:w(s)_Caprini}
\omega(s)=\frac{\sqrt{s}-\alpha \sqrt{4\,m_K^2-s}}{\sqrt{s}+\alpha \sqrt{4\,m_K^2-s}}\,,
\end{equation}
with the typical choice of $\alpha=1.0$ \cite{Caprini:2008fc}. The variation of $\alpha$ around $1.0$ or replacing Eq.(\ref{Eq:w(s)_Caprini}) by (\ref{xi-1}), leads to the very compatible results. The two parameter fit to Roy solution up to \mbox{$s_{max}^{1/2}=700$} MeV gives $\chi^2/\text{d.o.f}=1.7$,~\footnote{The fit of Eqs.(\ref{Eq:Caprini},\ref{Eq:w(s)_Caprini}) to Roy-like solution up to \mbox{$s_{max}^{1/2}=800$} MeV gives $\chi^2/\text{d.o.f}=4.0$ compared to $\chi^2/\text{d.o.f}=1.5$ using Eq.(\ref{DRfor1/T_finalJ0}).} and the pole position at \mbox{$\sqrt{s_p}=448(7)-i\,205(4)\,\text{MeV}$}. In order to improve the obtained width of $\sigma/f_0(500)$, one needs to add a third parameter. This is clearly not as good as the fit given in Table \ref{tab:FitResults}, which produces a pole position at \mbox{$\sqrt{s_p}=468(8)-i\,239(4)\,\text{MeV}$}, i.e. much closer to the Roy solution with just two parameters. Another useful comparison of Eq.(\ref{DRfor1/T_finalJ0}) and Eqs.(\ref{Eq:KmatrixGeneral},\ref{Eq:Caprini},\ref{Eq:w(s)_Caprini}) is the employment of them as the unitarization method of $\chi$PT. By constraining two unknown parameters in Eqs.(\ref{Eq:KmatrixGeneral},\ref{Eq:Caprini},\ref{Eq:w(s)_Caprini}) and Eq.(\ref{DRfor1/T_finalJ0}) from the $\chi$PT threshold parameters $a_{0, \text{NNLO}}=0.220(5)$ and $b_{0, \text{NNLO}}=0.276(6)$ \cite{Colangelo:2001df}
one obtains $\sqrt{s_p}=389(15)-i\,262(13)\,\text{MeV}$ and $\sqrt{s_p}=426(29)-i\,263(22)\,\text{MeV}$, respectively. Clearly, the dispersive parameterization (\ref{DRfor1/T_finalJ0}) produces the results closer to the Roy-like analysis.

For the description of the lattice data of S-wave isoscalar $\pi\pi\to \pi\pi$ ($m_\pi=239$ MeV) scattering \cite{Briceno:2016mjc} HadSpec collaboration used 8 different parameterizations, leading to the wide spread of the $\sigma/f_0(500)$ pole position, $\sqrt{s_p}=550 \ldots 780-i\,(115 \ldots 285)\,\text{MeV}$. We would like to emphasize, that out of all applied parameterizations, only the "fit 3a" (which corresponds to Eqs.(\ref{Eq:KmatrixGeneral},\ref{Eq:Kmatrix2}) with $\gamma_n=0$ and $I(s)=-\text{Re}[ R(m^2,s_{th})]+R(s,s_{th})$) can be cast into the dispersive form for the inverse amplitude with the conformal series truncated at $n_{max}=0$. However, it translates one-to-one
only under an assumption that $g^2$ and $m^2$ in Eq.(\ref{Eq:Kmatrix2}) can also take on negative values. It can also be reinforced by the fact that in order to reproduce $\chi$PT threshold parameters for the physical pion mass $a_{0, \text{NNLO}}=0.220$ and $b_{0, \text{NNLO}}=0.276$, one needs to use $g^2=-3.78$ and $m^2=-1.47$. In turn, for $m_\pi=239$ MeV the "fit 3a" from HadSpec and Eq.(\ref{DRfor1/T_finalJ0}) truncated at $n_{max}=0$ coincide, leading to a significant reduction of the spread of the $\sigma/f_0(500)$ pole position. When more precise data will be available, the extension of "fit 3a" dispersively goes with additional parameters related to the left-hand cuts $C_{n}$ in Eqs.(\ref{DRfor1/T_finalJ0},\ref{xi-1}), rather than polynomial $\gamma_n$ terms in Eq.(\ref{Eq:Kmatrix2}). In Table \ref{tab:FitResults} we reproduced the results of "fit 3a" close enough, given the fact that we performed a simple fit to $p\,\cot\delta$ instead of the energy levels. The former is sufficient for our following discussion related to the chiral extrapolation.

As a first step, we quantify the uncertainty due to the Adler zero position. At NLO using LECs from \cite{Bijnens:2014lea} we obtain (see Fig.~\ref{Fig:sA})
\begin{equation}
    s^{I=0}_{A,\text{NLO}}(m_\pi = 239\,\mathrm{MeV})= 0.023(10)\, \mathrm{GeV}^2\,,
\label{AdlerZeroNLOsu2}
\end{equation}
which is a very conservative estimate, since the LECs were assumed to be uncorrelated. The propagation of Eq.(\ref{AdlerZeroNLOsu2}) into the pole position gives $\sqrt{s_p} = 561(4)-i\,171(7)\,\mathrm{MeV}\,$. In case when both the uncertainties of lattice data and Adler zero input are taken into account, one gets
\begin{equation}
\sqrt{s_p} = 559^{+48}_{-53}-i\,168^{+20}_{-17} \,\mathrm{MeV}\,,
\end{equation}
where the error corresponds to $1\sigma$ confidence level provided by the bootstrap analysis. We have also checked that exactly the same result is achieved when only the uncertainty of lattice data is accounted for, while the Adler zero is fixed to its central NLO position. This points out that the current lattice data completely dominate the uncertainty, and the final result is not very sensitive to the exact position of the Adler zero in Eq.(\ref{AdlerZeroNLOsu2}). 

Next, we compare the values of the fitted parameters to lattice data \cite{Briceno:2016mjc}
\begin{align}\label{Eq:Parameters}
    g_{A} &= 0.58(10),\quad C_{0} = 1.36(26)\,,
\end{align}
with the corresponding estimations from NLO $\chi$PT ($m_\pi=239$ MeV)
\begin{align}\label{Eq:Parameters}
g_{A, \text{NLO}}&=\left(\left.\frac{\text{d} t^{\text{NLO}}_0(s)}{\text{d} s}\right\vert_{s=s_A}\right)^{-1}= 0.45(3)\,,\nonumber \\
C_{0,\text{NLO}}&=\left[t^{\text{NLO}}_0(s_{th})\right]^{-1}= 1.42(6)\,.
\end{align}

As one can see, the fitted parameters are consistent with $\chi$PT extrapolation. Unfortunately, currently the error bars of the lattice data are too large to impose some constraints on the LECs. However, more data is expected in the near future with the improved precision. In addition, the existing lattice data for the P-wave \cite{Dudek:2012xn,*Wilson:2015dqa,Bruno:2016plf} (calculated also at the physical pion mass \cite{Fischer:2020yvw,Paul:2021pjz}) can help to constraint LECs even at the two loop level, as has been demonstrated in \cite{Niehus:2020gmf}. We emphasise that in the proposed dispersive parametrizations, 
we are free to choose where to do the matching to $\chi$PT. The natural choice is around the Adler zero and/or the threshold.
Once the matching is done, the proposed parameterizaition can be used to predict the pion mass dependence 
and the obtained LECs will correspond exactly to the ones in perturbative $\chi$PT calculations, as opposed to mIAM (IAM). We have checked that Eq.(\ref{DRfor1/T_finalJ0}), truncated to the leading term in the conformal expansion with parameters fixed from NLO $\chi$PT (see left hand side of Eq.(\ref{Eq:Parameters})), produces the same qualitative behaviour of the $f_0(500)$ pole as in mIAM \cite{Hanhart:2008mx}. With increasing pion mass values the imaginary part of the pole decreases, and then $f_0(500)$ becomes a virtual bound state.

\section{Conclusion and outlook}
\label{sec:Conclusion and outlook}

In this work, we presented improved parametrizations for elastic p.w. amplitudes, see Eqs. (\ref{DRfor1/T_finalJ0},\,\ref{DRfor1/T_finalJ1_m1=m2},\,\ref{DRfor1/T_finalJ2_m1=m2},\,\ref{DRfor1/T_finalJ1_m1!=m2}), which are based on dispersive representations for the inverse amplitudes. In this approach unitarity and analyticity constraints are implemented exactly. The contributions from the left-hand cuts were accounted for in a model-independent way using the expansion in a conformal variable, which maps the left-hand cut plane onto the unit circle. For the S-wave scattering special attention was paid to the possible Adler zero contribution. For the higher partial waves we implemented carefully the angular momentum barrier factors. We also compared our approach with the mIAM (IAM) and argued that both for the S and P-waves the constructed parametrizations can be understood as a more general method, where one is not assuming a particular Lagrangian-based form for the left-hand cuts and subtraction constants. The latter is useful for lattice calculation with the relatively large $m_\pi$ or scattering of $\pi K$, $\pi D$, $K D$, etc.

We applied the new parametrizations to the well-studied test cases of $\pi\pi$ and $\pi K$ scattering with $J=0,1,2$, showing that at most 3 parameters are needed to reproduce very precise Roy/Roy-Steiner pseudo-data in the physical region. The obtained pole positions lie fairly close to the exact solutions. 

The main motivation for developing these amplitudes, which we call dispersive inverse amplitudes (DIA), is their application to the upcoming lattice data in $\pi\pi$, $\pi K$, $\pi D$ and $K D$ channels. In addition, these parametrizations can be beneficial for the fits to data constrained by Roy-like equations or forward dispersion
relations \cite{Pelaez:2015qba,GarciaMartin:2011cn,*GarciaMartin:2011jx,*Pelaez:2019eqa,Pelaez:2016tgi,Pelaez:2018qny}. However, the framework is general and not specific to the particular reactions, thus laying the groundwork for analyses of any lattice or experimental data in the elastic region.

The proposed dispersive amplitudes are written in  compact analytical forms, which are well-suited for direct numerical implementations. A Mathematica file with these formulas is provided as supplemental material.

\section*{Acknowledgements}
I.D. acknowledges useful discussions with Cesar Fern\'andez-Ram\'irez. This work was supported by the Deutsche Forschungsgemeinschaft (DFG, German Research Foundation), in part through the Research Unit [Photon-photon interactions in the Standard Model and beyond, Projektnummer 458854507 - FOR 5327], and in part through the Cluster of Excellence [Precision Physics, Fundamental Interactions, and Structure of Matter] (PRISMA$^+$ EXC 2118/1) within the German Excellence Strategy (Project ID 39083149).

\appendix

\section{Analytic form of $R(s,\tilde{s}_M)$ }
\label{Appendix:R}
Here we collect the analytic expressions for 
\begin{align}\label{R(s,s_th)}
R(s,\tilde{s}_M)\equiv & \frac{s-\tilde{s}_{M}}{\pi} \int_{s_{th}}^{\infty}\frac{d s'}{s'-\tilde{s}_{M}}\frac{-\rho(s')}{s'-s}\,.
\end{align}
For $\tilde{s}_M=s_{th}$ it holds \cite{Wilson:2014cna} (see also \cite{Basdevant:1977ya,*Oller:1998zr})
\begin{align}\label{R(s,s_th)_analyt}
R(s,s_{th})&=\frac{\rho(s)}{\pi}\log \left[ \frac{\xi(s)+\rho(s)}{\xi(s)-\rho(s)} \right]-
\frac{\xi(s)}{\pi}\frac{m_2-m_1}{m_2+m_1}\log \frac{m_2}{m_1},\nonumber\\
 \xi(s) &\equiv 1-\frac{s_{th}}{s} \\
&\overset{m_1=m_2} {=}-\frac{\rho(s)}{\pi}\log \left[ \frac{\rho(s)-1}{\rho(s)+1} \right]\,, \nonumber
\end{align}
while for any $\tilde{s}_M<s_{th}$ it is given by
\begin{align}\label{R(s,s_M)analytic}
R(s,\tilde{s}_M)&=R(s,s_{th})-R(\tilde{s}_M,s_{th})\,.
\end{align}
Note that for $\tilde{s}_M=0$, the function $R(s,0)$ is nothing else, but the well-known one-loop function \cite{GomezNicola:2001as}
\begin{align}\label{R(s,0)}
R(s,0)&=R(s,s_{th})-R(0,s_{th})\nonumber\\
&=R(s,s_{th})-\frac{1}{\pi}\left(1+\frac{2\,m_1\, m_2}{m_2^2-m_1^2}\,\log \frac{m_2}{m_1}\right)\\
&=-16\,\pi\,\bar{J}(s)\,.\nonumber
\end{align}
The numerical implementation of Eq.~(\ref{R(s,s_th)_analyt}) has to be performed with an extra care, since it has to give the correct result not only in the physical region (see Fig.~\ref{Fig:R(s,sth)}), but also in the complex plane. In the supplemental material we provide the results for $R(s,\tilde{s}_M)$ in Mathematica.

\begin{figure}[h]
\centering
\includegraphics[width =0.40\textwidth]{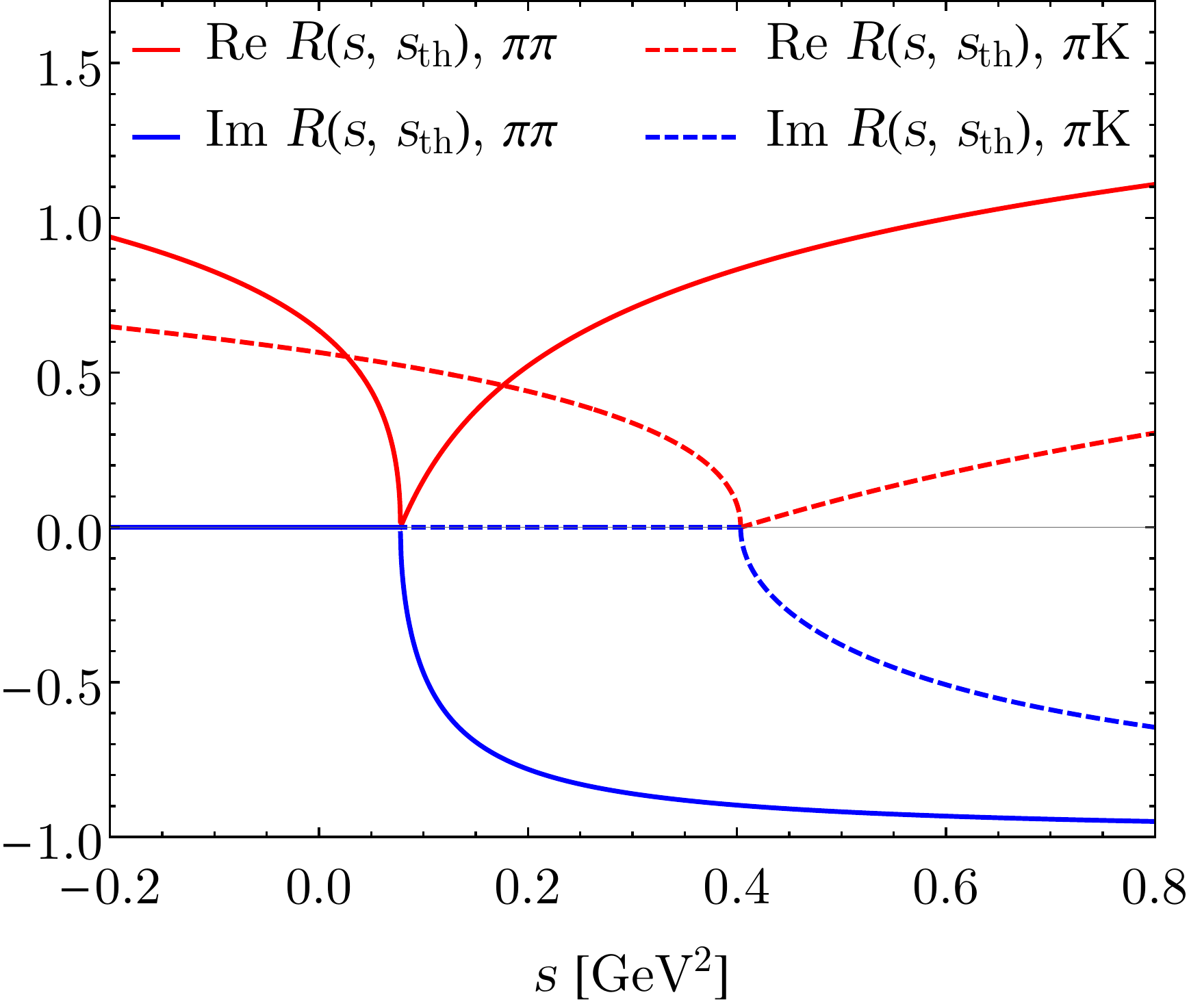}
\caption{Real and imaginary parts of $R(s,s_{th})$ for $\pi\pi \to \pi\pi$ and $\pi K \to \pi K$ scattering.}
\label{Fig:R(s,sth)}
\end{figure}

\section{$J \neq 0$, $m_1 \neq m_2$ scattering}
\label{Appendix:B}
In this appendix, we show the general derivation of the DIA for $J \neq 0$, $m_1 \neq m_2$. We propose to write a $J+1$ subtracted spectral representation for
\begin{equation}\label{g_J_2}
g_J(s)\equiv \frac{p(s)^{2J}}{t_J(s)}\,,
\end{equation}
which is free from any kinematic constraints. It holds
\begin{align}\label{DR_g_J}
g_J(s)=P_J(s)+\frac{Q_{J+1}(s)}{\pi} \int_{L,R}\frac{d s'}{Q_{J+1}(s')}\frac{\text{Im } g_J(s')}{s'-s},
\end{align}
where $P_J(s)$ is a polynomial of degree $J$ and $Q_{J+1}(s)$ is defined as
\begin{align}\label{g_J}
Q_{J+1}(s)\equiv (s-s_{M_1})(s-s_{M_2})\,...\,(s-s_{M_{J+1}})\,.
\end{align}
The choice of the subtraction points $s_{M_i}$ is in general arbitrary. However, it is useful to choose $s_{M_i}$ in such a way, that the integral over the right-hand cut can be written in terms of $R(s,\tilde{s}_M)$ given in Eq.(\ref{R(s,s_th)}) (which is known analytically). Therefore, for the case of $m_1=m_2$ it is useful to put all subtraction points at the threshold $s_{M_i}=s_{th}$, thus reproducing Eq.(\ref{DRfor1/T_J1_0}). On another side, for $J=1$ and $m_1 \neq m_2$, it is useful to choose $s_{M_1}=m_-^2$ and $s_{M_2}=m_+^2$, which leads to Eq.(\ref{DRfor1/T_finalJ1_m1!=m2}). For $J>1$ and $m_1 \neq m_2$ it is unfortunately impossible to express the answer in terms of $R(s,\tilde{s}_M)$. For simplicity, let us choose $s_{M_i}$ at the same point $s_{M}$
\begin{align}\label{DRfor1/TJ_m1!=m2_0}
&\left[t_J(s)\right]^{-1}=\frac{1}{p^{2J}(s)}
\left(\sum_{i=0}^{J}\frac{1}{i
!}\,g_J^{(i)}(s_M)\,(s-s_M)^i\right.\\
&\left.+\frac{(s-s_M)^{J+1}}{\pi} \int_{L,R}\frac{d s'}{s'-s_M}\frac{\text{Im } \left[t_J(s)\right]^{-1}}{s'-s}\frac{p^{2J}(s')}{(s'-s_M)^J}\right)\nonumber
\end{align}
and approximate (as described before) the contribution from the left-hand cut together with the constant term in a suitably constructed conformal mapping series. It holds
\begin{align}\label{DRfor1/TJ_m1!=m2_1}
\left[t_J(s)\right]^{-1}=&\frac{1}{p^{2J}(s)}
\left(\sum_{i=0}^{J-1}a_i\,(s-s_M)^i\right.\\
&\left.\quad +(s-s_M)^{J}\left\{\sum_{n=0}^\infty C_{n}\,\omega^n(s)+R_J(s,s_M)\right\}\right)\nonumber\,,
\end{align}
where
\begin{align}
R_J(s,s_M)\equiv \frac{s-s_M}{\pi} \int_{s_{th}}^{\infty}\frac{d s'}{s'-s_M}\frac{-\rho(s')}{s'-s}\frac{p^{2J}(s')}{(s'-s_M)^J}\,.
\end{align}
In Eq.(\ref{DRfor1/TJ_m1!=m2_1}), the constants $a_i$ and $C_n$ are unknown.

\bibliographystyle{apsrevM}
\bibliography{PhysLettB}

\end{document}